\documentclass[12pt,english]{amsart}
\usepackage{mathpazo}
\usepackage{multirow}
\usepackage{helvet}
\usepackage{courier}
\usepackage[T1]{fontenc}
\usepackage[latin9]{inputenc}
\usepackage[letterpaper]{geometry}
\usepackage{babel}
\usepackage{units}
\usepackage{amsthm}
\usepackage{amssymb}
\usepackage{esint}
\PassOptionsToPackage{normalem}{ulem}
\usepackage{ulem}
\usepackage[unicode=true,
 bookmarks=true,bookmarksnumbered=false,bookmarksopen=false,
 breaklinks=false,pdfborder={0 0 1},backref=false,colorlinks=false]
 {hyperref}
\hypersetup{
 pdfauthor={Jeffrey Schenker}}

\makeatletter

\providecommand{\tabularnewline}{\\}

\numberwithin{equation}{section}
\numberwithin{figure}{section}
\theoremstyle{plain}
\newtheorem{thm}{\protect\theoremname}
  \theoremstyle{remark}
  \newtheorem*{rem*}{\protect\remarkname}

\newcommand{\ip}[1]{\left\langle #1 \right \rangle}
\newcommand{\ipc}[2]{\left \langle #1 \, , \, #2 \right \rangle}
\newcommand{\Ev}[1]{\E \left( #1 \right)}  %% produces \E( # )

\newcommand{\abs}[1]{\left\vert#1\right\vert}
\newcommand{\set}[1]{\left\{#1\right\}}

\def\e{\mathrm e}

\def\half {\frac{1}{2}}
\def\1{{\mathsf 1}}
\def\di{\mathrm d}
\def\Im{\operatorname{Im}}
\def\Z{\mathbb Z}

\def\R{\mathbb R}
\def\C{\mathbb C}
\def\E{\mathbb E}

\makeatletter
\def\rightharpoondownfill@{%
    \arrowfill@\relbar\relbar\rightharpoondown}
\def\rightharpoonupfill@{%
    \arrowfill@\relbar\relbar\rightharpoonup}
\def\leftharpoondownfill@{%
    \arrowfill@\leftharpoondown\relbar\relbar}
\def\leftharpoonupfill@{%
    \arrowfill@\leftharpoonup\relbar\relbar}
\newcommand{\xrightharpoondown}[2][]{%
    \ext@arrow 0359\rightharpoondownfill@{#1}{#2}}
\newcommand{\xrightharpoonup}[2][]{%
    \ext@arrow 0359\rightharpoonupfill@{#1}{#2}}
\newcommand{\xleftharpoondown}[2][]{%
    \ext@arrow 3095\leftharpoondownfill@{#1}{#2}}
\newcommand{\xleftharpoonup}[2][]{%
    \ext@arrow 3095\leftharpoonupfill@{#1}{#2}}
\newcommand{\xleftrightharpoons}[2][]{\mathrel{%
    \raise.22ex\hbox{%
        $\ext@arrow 3095\leftharpoonupfill@{\phantom{#1}}{#2}$}%
    \setbox0=\hbox{%
        $\ext@arrow 0359\rightharpoondownfill@{#1}{\phantom{#2}}$}%
    \kern-\wd0 \lower.22ex\box0}%
}
\newcommand{\xrightleftharpoons}[2][]{\mathrel{%
    \raise.22ex\hbox{%
        $\ext@arrow 3095\rightharpoonupfill@{\phantom{#1}}{#2}$}%
    \setbox0=\hbox{%
        $\ext@arrow 0359\leftharpoondownfill@{#1}{\phantom{#2}}$}%
    \kern-\wd0 \lower.22ex\box0}%
} \makeatother

\makeatother

  \providecommand{\remarkname}{Remark}
\providecommand{\theoremname}{Theorem}

\begin{document}

\title[How large is large?]{How large is large? \\
\small Estimating the critical disorder for the Anderson model}

\author{Jeffrey Schenker}
\address{Michigan State University}
\email{jeffrey@math.msu.edu}
\thanks{Supported by NSF CAREER Award DMS-08446325.}
\begin{abstract} Complete localization is shown to hold for the $d$-dimensional Anderson model with uniformly distributed random potentials provided the disorder strength $\lambda >\lambda_{And}$ where $\lambda_{\text{And}}$ satisfies $\lambda_{\text{And}}=\mu_d e \ln \lambda_{\text{And}}$ with $\mu_d$ the self-avoiding walk connective constant for the lattice $\Z^d$.  Notably $\lambda_{\text{And}}$ is precisely the large disorder threshold proposed by Anderson in 1958.
\end{abstract}

\date{May 29, 2013; revised Oct. 17, 2014}

\maketitle

\section{Introduction}
The aim of this note is to present an improved rigorous estimate of the critical disorder strength for the onset of complete localization of the eigenstates of the Anderson tight binding Hamiltonian on $\ell^{2}(\mathbb{Z}^{d})$, 
\begin{equation}
H_{\omega}\psi(x)\ =\ \sum_{|x'-x|=1}\psi(x')+\lambda\omega(x)\psi(x),
\end{equation}
where $\omega(x)$ are mutually independent random variables uniformly distributed in $[-1,1]$. The number $\lambda$ is called
the \emph{disorder strength.} 

This Hamiltonian was first studied by Anderson \cite{Anderson:1958p1895} who proposed that at sufficiently large disorder the states are localized, resulting in dynamical trapping of any wave packet.  Specifically,  it was proposed that complete localization occurs for $\lambda >\lambda_{\text{And}}$ where $\lambda_{\text{And}}$ is the unique solution of the equation
\begin{equation}\label{eq:lambdacrit}
\lambda_{\text{And}}=\mu_d e \ln \lambda_{\text{And}}.
\end{equation}
Here $\mu_d$ is the connectivitiy constant for self-avoiding walks on $\Z^d$ and $e$ is the base of the natural logarithm.  (See \cite[Eq.\ (84)]{Anderson:1958p1895}. In the notation of \cite{Anderson:1958p1895}, $\lambda=\nicefrac{W}{2V}$ and $\mu_d=K$.)

The methods employed in \cite{Anderson:1958p1895} were heuristic and involved uncontrolled approximations, however the Hamiltonian $H_\omega$ is well studied in the mathematical literature. The earliest proof of localization for dimensions $d>1$ is due to Fr\"ohlich, Spencer, Martinelli and Scopola \cite{Frohlich:1985vx} based on the multi-scale analysis of \cite{Frohlich:1983vs}. In \cite{Frohlich:1985vx} complete localization is shown to hold for \emph{$\lambda$ sufficiently large}, but no concrete estimate of the critical disorder is given. Later Aizenman-Molchanov \cite{Aizenman:1993vs} introduced a simplified "moment-method" for proving localization, based on which it is easier to obtain a specific estimate of critical disorder.  For instance Aizenman and Graf \cite{Aizenman:1998wx} showed that complete localization holds for $H_\omega$ provided
\begin{equation}\label{eq:AG}
\lambda^{s}>\frac{4d}{1-s}
\end{equation}
for some $s<1$.  Optimizing over $s$ one finds that localization holds provided $\lambda>\lambda_{\text{AG}}$ where 
\begin{equation}\label{eq:lambdaAG}
\lambda_{\text{AG}} = 4d e \ln \lambda_{\text{AG}}.
\end{equation}
Since the connectivity $\mu_d < 2d-1$, $\lambda_{\text{AG}}> \lambda_{\mathrm{And}}$, i.e., this estimate of critical disorder is weaker than Anderson's --- in fact it is quite a bit weaker; see Table \ref{tab:values}. 

The main result of this note is that complete localization holds for $H_\omega$ throughout the region $\lambda >\lambda_{\text{And}}$ proposed in \cite{Anderson:1958p1895}. In dimension $d=3$ the best known upper bound on $\mu_d$ is $\mu_3 \le 4.7114$ \cite[Table 5.2]{finch2003mathematical}, leading to a critical disorder of $\lambda_{\text{And}}\le 50.3$, as compared to the value $\lambda_{\text{AG}}\approx 167$.  See Table \ref{tab:values} for a more detailed comparison in dimensions $2$ to $6$.

\begin{table}
\begin{tabular}{r|ccccc}
& \multicolumn{5}{|c}{Dimension} \tabularnewline
&$2$  & $3$ & $4$ & $5$  & $6$ \tabularnewline
\hline
$\mu_{d}^{*}$ & 2.68 & 4.72 & 6.81 & 8.86 &  10.89  \tabularnewline 
$\lambda_{\text{And}}^*$ &   22.8  & 50.3 & 81.7  & 114.1 &  148.0  \tabularnewline
$\lambda_{AG}$ & 100.2 & 167.0   &   238.1 & 312.3 & 389.1
\end{tabular}
\vskip .1in
\caption{\label{tab:values}  For $d=2,3,4,5$, and $6$, the table shows: 1)  The best rigorous upper bound $\mu_d^*$   for $\mu_d$ according to \cite[Table 5.2]{finch2003mathematical}; 2) The corresponding root $\lambda_{\text{And}}^*$  of \eqref{eq:lambdacrit} with $\mu_d^*$ in place of $\mu_d$, rounded up in the last digit so that Theorem \ref{thm:main} guarantees localization for $\lambda >\lambda_{\text{And}}^*$; and  3) the Aizenman-Graf critical disorder 
$\lambda_{\text{AG}}$ satisfying \eqref{eq:lambdaAG}. }
\end{table}

The estimates obtained here are based on a modified version of the Aizenman-Molchanov method \cite{Aizenman:1993vs}, with improvements resulting from 1) avoiding the ``decoupling estimate'' used in \cite{Aizenman:1993vs, Aizenman:1998wx}  and 2) incorporating known facts about self-avoid walks. Neither of these improvements is particularly novel.  Decoupling estimates were avoided in \cite{Aizenman:2001vz} using a two-step perturbation argument --- which is, however, inefficient in the large disorder regime discussed here.  Similarly, the connection between localization and self-avoiding walks has been known since Anderson's original work \cite{Anderson:1958p1895}. However, the observation that Anderson's original heuristic estimate of critical disorder is the best estimate provable by current methods appears to be new.

Self-avoiding walks have also appeared in mathematical studies of large disorder localization in a survey by Hundertmark \cite{Hundertmark:2008gx} and more recently in works of Tautenham \cite{Tautenhahn:2011iv} and Suzuki \cite{Suzuki:2013gz}.  The work of Tautenhahn \cite{Tautenhahn:2011iv}, in particular, is quite closely related to the present work.  He studied localization on arbitrary locally finite graphs and obtained a result \cite[Theorem 2.2]{Tautenhahn:2011iv} bounding the resolvent at large disorder in terms of the connectivity of self-avoiding walks for sufficiently large disorder. In fact, the main result proved here (Theorem \ref{thm:main}) could be obtained from \cite[Theorem 2.2]{Tautenhahn:2011iv}, specialized to $\Z^d$, via a short optimization argument.  However, to keep the present paper self contained and to facilitate the discussion of different methods of proving large disorder localization, a self contained proof of Theorem \ref{thm:main} is presented below.

\section{Self-avoiding walks and localization}
A \emph{self-avoiding walk in $\mathbb{Z}^{d}$} of length
$N$ is an ordered $N$-tuple of \emph{distinct} points $x_{0},x_{1},\ldots,$
$x_{N}\in\mathbb{Z}^{d}$ such that $\left|x_{i}-x_{i-1}\right|=1$
for $i=1,\ldots,N$. Let us denote the set of all self-avoiding walks of length $N$ starting at $x_{0}=x$ and ending at $x_{N}=y$ by
$\mathcal{S}_{N}\left(y,x\right)$. The \emph{self-avoiding walk correlation function}
is 
\begin{equation}\label{eq:Ct}
C_{\gamma}\left(y-x\right):=\sum_{N=0}^{\infty}\gamma^{N}\#\mathcal{S}_{N}\left(y,x\right),
\end{equation}
defined at those values of $\gamma$ for which the right hand side is absolutely summable. As indicated the correlation function
depends only on the difference $y-x$, a fact which is made obvious
by noting that walks from $x$ to $y$ are in one-to-one correspondence
with walks from $0$ to $y-x$. The \emph{self-avoiding walk susceptibility} is the
sum of the correlation function
\begin{equation}
\chi\left(\gamma\right):=\sum_{x}C_{\gamma}\left(x\right)=\sum_{N=0}^{\infty}c_{N}\gamma^{N}\label{eq:susceptibility}
\end{equation}
where $c_{N}$ denotes the number of all self-avoiding walks of length
$N$ starting at $0$, $c_{N}=\sum_{x}\#\mathcal{S}_{N}\left(x,0\right)$.
The following summarizes known properties of the susceptibility and correlation function (see \cite[Chapter 1]{neal1996self}):
\begin{enumerate}
\item The connectivity constant of $\Z^d$, which is the limit
\[
\mu_{d}:=\lim_{N\rightarrow\infty}\left(c_{N}\right)^{\nicefrac{1}{N}},
\]
exists, is positive and is bounded by $2d-1$. As a consequence the
power series in \eqref{eq:susceptibility} has a finite, positive
radius of convergence $\nicefrac{1}{\mu_{d}}$. 
%\item When$0\le\gamma <\nicefrac{1}{\mu_d}$ we have
%\[
%\chi\left(\gamma\right)\ge\frac{1}{1-\mu_{d}\gamma }.
%\]
%In particular 
%\[
%\lim_{\gamma \uparrow\nicefrac{1}{\mu_{d}}}\chi\left(\gamma \right)=\infty.
%\]

\item When $0\le\gamma <\nicefrac{1}{\mu_d}$ the correlation function $C_{\gamma }\left(x\right)$
decays exponentially as $x\rightarrow\infty$. Specifically, for
any $\epsilon>0$ there is $K_{\epsilon}<\infty$ such that 
\[
C_{\gamma}\left(x\right)\le K_{\epsilon}\left(\left(\mu_{d} +\epsilon\right) \gamma  \right)^{|x|}
\]
where $\left|x\right|$ denotes the $\ell^{1}$ norm of a lattice
vector: $|x|=\left|x_{1}\right|+\cdots\left|x_{d}\right|.$
\end{enumerate}

Our main result for the Anderson Hamiltonian $H_\omega$ is conveniently stated in terms
of the Green's functions of the restrictions of $H_{\omega}$ to various
subsets of $\mathbb{Z}^{d}$. Let $\Lambda\subset\mathbb{Z}^{d}$
be any subset, infinite or finite, and define 
\begin{equation}
H_{\omega}^{(\Lambda)}\psi(x)\ =\ \sum_{\substack{|x'-x|=1\\
x'\in\Lambda
}
}\psi(x')+\lambda\omega(x)\psi(x),
\end{equation}
for $\psi \in \ell^2(\Lambda)$,
and the associated Green's function 
\begin{equation}
G_{z}^{(\Lambda)}(x,y)\ =\ \begin{cases}
\ip{\delta_{x},(H_{\omega}^{(\Lambda)}-z)^{-1}\delta_{y}}, & \mbox{if }x,y\in\Lambda\\
0\ , & \mbox{ if }x\in\Lambda^{c}\mbox{ or }y\in\Lambda^{c}.
\end{cases}
\end{equation}
For any $\Lambda$, the Green's function is well defined for $z\in\C\setminus[-2d-\lambda,2d+\lambda]$.
For finite $\Lambda$ it also makes sense (almost surely)
for $z\in[-2d-\lambda,2d+\lambda]$. 
\begin{thm}\label{thm:main}
Let $\gamma(\lambda)=\frac{e\ln\lambda}{\lambda}$, where $e$ is
the base of the natural logarithm. If $\lambda>e$ and $\mu_{d}\gamma(\lambda)< 1$
then 
\begin{equation}
\Ev{\abs{G_{z}^{(\Lambda)}(x,y)}^{1-\frac{1}{\ln\lambda}}}\le\left(\ln\lambda\right)C_{\gamma(\lambda)}\left(x-y\right)
\end{equation}
for any subset $\Lambda\subset\mathbb{Z}^{d}$ and for all $z\in\C\setminus\left[-2d-\lambda,2d+\lambda\right]$.
In particular, if $\lambda_{\mathrm{And}}$ satisfies \eqref{eq:lambdacrit} then for all $\lambda>\lambda_{\mathrm{And}}$ and $\epsilon>0$ we have
\begin{equation}\label{eq:specificFME}
\Ev{\abs{G_{z}^{(\Lambda)}(x,y)}^{1-\frac{1}{\ln\lambda}}}\le K_{\epsilon}\ln\lambda e^{-m_{\epsilon}(\lambda)|x-y|}
\end{equation}
where 
\[
m_{\epsilon}(\lambda)=-\ln\gamma(\lambda)-\ln\left(\mu_{d}+\epsilon\right)>0.
\]
 \end{thm}
\begin{rem*} It follows from H\"older's inequality that whenever $\lambda>\lambda_{\text{And}}$
we have for each $s\in(0,1)$ constants $C_{s}<\infty$ and $\mu_{s}>0$
such that 
\begin{equation}\label{eq:FME}
\Ev{\abs{G_{z}^{(\Lambda)}(x,y)}^{s}}\le C_{s}\e^{-\mu_{s}|x-y|}.
\end{equation} 
Eq. \eqref{eq:FME} implies exponential dynamical localization via the following estimate for the propagator
$$ \Ev{\sup_{t \in \R} \abs{\ip{\delta_x, e^{-i tH_\omega} \delta_y}}} \le A e^{-\mu |x-y|}$$ with $A<\infty$ and $\mu >0$.  This is proved in \cite[Appendix B]{Aizenman:2001vz}.  
\end{rem*}

\section{Moment bounds and decoupling estimates}
The  Aizenman-Molchanov moment method \cite{Aizenman:1993vs, Aizenman:1994p8097, Aizenman:1998wx} relies on  two elementary estimates, valid for a fractional exponent $s<1$,
\begin{enumerate}
\item the fractional moment bound, 
\begin{equation}\label{eq:FMB}
\half\int_{-1}^{1}\frac{1}{\abs{V-B}^{s}}\di V\ \le\ \frac{1}{1-s}\quad\mbox{ for all }B\in\mathbb{C},
\end{equation}
which is clearly saturated for $B=0$; and
\item 
the ``decoupling estimate'' 
\begin{equation}
\frac{\int_{-1}^{1}\frac{|V-A|^{s}}{|V-B|^{s}}\di V}{\int_{-1}^{1}\frac{1}{|V-B|^{s}}\di V}\ge D(s)>0,\quad\mbox{for all }A,B\in\mathbb{C},\label{eq:decouple}
\end{equation}
with $D(s)$>0. 
\end{enumerate}

In \cite{Aizenman:1998wx} complete localization is obtained provided 
$$ \lambda^s >  \frac{2d}{ D(s)} .$$
The symmetries of \eqref{eq:decouple} strongly suggest that the left hand side of \eqref{eq:decouple} is minimized for $A=B=0$, suggesting that the optimal value of $D(s)$ is $1-s$; however whether this is correct or not seems not to be known. Instead, in \cite{Aizenman:1998wx} it was shown that \eqref{eq:decouple} holds with 
\begin{equation}
D(s)=\frac{1-s}{2},
\end{equation}
leading to the large disorder criterion \eqref{eq:AG}.

It turns out that one may simply avoid the decoupling estimate by making use of the following ``depleted resolvent identity,'' 
\begin{equation}
G_{z}^{(\Lambda)}(x,y)=-G_{z}^{(\Lambda)}(x,x)\sum_{\substack{x'\in\Lambda\\
|x'-x|=1
}
}G^{(\Lambda\setminus\{x\})}(x',y),\label{eq:onestep}
\end{equation}
valid for $x\neq y$. To verify eq.\ \eqref{eq:onestep} we will apply the resolvent identity $$A^{-1}= B^{-1} - A^{-1} (A-B) B^{-1}$$ with $A= H^{(\Lambda)}-z$ and $B = H^{(\{x\})}\oplus H^{(\Lambda\setminus\{x\})} -z$, which is $A$ with all the hopping matrix elements from or to $x$ suppressed.  The result is
\begin{multline}\label{eq:resolvent} \left ( H^{(\Lambda)}- z\right )^{-1} \ = \ \left ( H^{(\{x\})}\oplus H^{(\Lambda\setminus\{x\})} -z \right )^{-1} \\
- \sum_{|x'-x|=1}\left ( H^{(\Lambda)}- z\right )^{-1} T_{x,x'} 
\left ( H^{(\{x\})}\oplus H^{(\Lambda\setminus\{x\})} -z \right )^{-1} 
\end{multline}
where $$T_{x,x'} \psi(y) \ = \ \begin{cases} \psi(x') & y=x \\ \psi(x) & y=x'\\
0 & y\not \in \{x,x'\}. 
 \end{cases}$$
Taking matrix elements now gives eq.\ \eqref{eq:onestep}.

If we apply eq.\ \eqref{eq:resolvent} to the factor of $(H^{(\Lambda)}-z)^{-1}$ that appears on the right hand side of that same equation and then take the $x,x$ matrix element of the resulting expression, we obtain the identity
\begin{multline*} G_z^{(\Lambda)}(x,x) \ = \ \frac{1}{\lambda \omega(x) -z} \\ - G_z^{(\Lambda)}(x,x) 
\sum_{|x''-x|=1} \sum_{|x'-x|=1} \ipc{\delta_{x''}}{    \left (  H^{(\Lambda\setminus\{x\})}- z\right )^{-1}  \delta_{x'}} \frac{1}{\lambda \omega(x)-z}. \end{multline*}
Solving for $G_z^{(\Lambda)}(x,x)$ yields
$$ G_z^{(\Lambda)}(x,x)=\frac{1}{\lambda \omega(x) -B(x,\omega)}$$
where $B(x,\omega)$ is a random variable independent of $\omega(x)$ (this identity also follows from the Schur formula for block matrix inversion). Thus the fractional moment bound \eqref{eq:FMB} gives the following
\emph{a priori} bound on the diagonal elements of the Green's function
\begin{equation}\label{eq:apriori}
\frac{1}{2}\int_{-1}^1 \abs{G_z^{(\Lambda)}(x,x)}^s  d \omega(x) \le \frac{1}{1-s}\frac{1}{\lambda^{s}}
\end{equation}
Since $\abs{ \sum_n a_n}^s \le  \sum_n |a_n|^s$, it follows that 
\begin{equation}\label{eq:DRB}
\half\int_{-1}^{1}|G_{z}^{(\Lambda)}(x,y)|^{s}d\omega(x)
\le\frac{1}{1-s}\frac{1}{\lambda^{s}}\sum_{\substack{x'\in\Lambda\\
|x'-x|=1
}
}\abs{G^{(\Lambda\setminus\{x\})}(x',y)}^{s}, \quad x\neq y
\end{equation}
 because the Green function $G^{(\Lambda\setminus\{x\})}$ is independent
of $\omega(x).$

At this point, if one is simply interested in a large disorder localization proof that avoids decoupling, but does not wish to optimize using self-avoiding walks, the following short argument will suffice.  Let 
\begin{equation}\label{eq:Fs}
F_{s}(x,y)=\sup_{\Lambda\subset\Z^{d}}\Ev{\abs{G_{z}^{(\Lambda)}(x,y)}^{s}}.
\end{equation}
Inserting a supremum over $\Lambda$ into \eqref{eq:DRB} and taking expectations we find 
\[
F_{s}(x,y)\le\frac{1}{1-s}\frac{1}{\lambda^{s}}\sum_{x'\ :\ |x'-x|=1}F_{s}(x',y),\quad x\neq y.
\]
If $$\Gamma_0(s) := \frac{2d}{1-s} \frac{1}{\lambda^s} < 1$$ it follows by iterating this argument multiple times along paths connecting $x$ to $y$ and using \eqref{eq:apriori} that
\begin{equation}\label{eq:loc!} F_{s}(x,y) \le \Gamma_0(s)^{|x-y|} F_{s}(y,y) \le \frac{1}{2d }  \Gamma_0(s)^{1+|x-y|}
\end{equation}
\emph{provided $F_{s}(x,y)$ is uniformly bounded as a function of $x$.}  

If we take $z\in \C\setminus \R$, then the necessary uniform bound is trivial since $F_{s}(x,y) \le \nicefrac{1}{|\Im z|^s} <\infty$. An analysis similar to that leading to \eqref{eq:FMB}, using a rank two Schur formula, can be used to show \emph{a priori} that $F_{s}(x,y)$ is bounded uniformly with regard to $z$ --- see \cite[Appendix B]{Aizenman:2001vz}.  However, we do not need this estimate as \eqref{eq:loc!} gives a uniform bound \emph{a posteriori}. 

Note that \eqref{eq:loc!} implies an estimate of the form \eqref{eq:FME} provided $\Gamma_0(s) <1$.  Thus to prove localization, for a given disorder strength $\lambda$ we must check if $\Gamma_0(s)<1$ for some $s<1$. To this end, note that $\Gamma_0(s)\rightarrow 2d$ 
as $s\rightarrow 0$.
For $\lambda <e$, one checks that $\Gamma_0(s)$ is strictly increasing, and thus never less than one.  However, for $\lambda >e$, there is a unique critical point 
\[
s_{\mathrm{crit}}=1-\frac{1}{\ln\lambda},
\]
in $(0,1)$. Since $\Gamma_0(s)\rightarrow \infty$ as $s\rightarrow 1$, we have 
\[
\min_{s\in (0,1)} \Gamma_0(s) = \Gamma_0(s_{\mathrm{crit}})=\frac{2de\ln(\lambda)}{\lambda},
\]
which is less than $1$ if 
\begin{equation}\label{eq:intermediate}
\lambda >2de \ln\lambda.
\end{equation}
This estimate already improves on \eqref{eq:lambdaAG}.  However, as we will see below, we can improve it even more using self-avoiding walks.

\section{Proof of Theorem \ref{thm:main}}
The above argument may be improved rather substantially if we iterate the depleted resolvent bound \eqref{eq:DRB} without maximizing over geometries.  

For instance, applying \eqref{eq:DRB} again to the resolvents sitting on the right hand side of that equation and averaging yields the estimate
\begin{multline*}
\mathbb{E}\left(\left|G^{(\Lambda)}(x,y)\right|^{s}\right)\le\frac{1}{\lambda^{2s}\left(1-s\right)^{2}}\sum_{\substack{x_1,x_2\in\Lambda\\ x_1\neq y, \
x_2 \neq x\\
|x_1-x|=|x_1-x_2|=1
}
}\mathbb{E}\left(\left|G^{(\Lambda\setminus\{x,x_1\})}(x_2,y)\right|^{s}\right) \\
+ I[|x-y|=1] \frac{1}{\lambda^{s}(1-s)} \Ev{\abs{G^{(\Lambda \setminus \{x\})}(y,y)}^s}.
\end{multline*}
Here $I[|x-y|=1] =1$ if $x$ and $y$ are neighbors and is zero otherwise;  the second term accounts for the fact that \eqref{eq:DRB} cannot be applied to a term of the form $G^{(\Lambda \setminus \{x\})}(y,y)$.
Since there are $2d(2d-1)$ choices for the points $x_1,x_2$ this yields the localization criterion 
\[
\lambda \ge\sqrt{2d(2d-1)}e\ \ln\lambda
\]
which improves on \eqref{eq:intermediate}.

However, we need not stop at two iterations. Suppose we apply \eqref{eq:DRB} $N$ times. The resulting sequences of points $x,x_1,\ldots,x_N$ depleted from the region $\Lambda$ make a self-avoiding walk. 
After averaging, we obtain the following estimate
\begin{multline*}
\mathbb{E}\left(\left|G^{(\Lambda)}(x,y)\right|^{s}\right) \le  \sum_{n=0}^N \left [ \frac{1}{\lambda^s (1-s)} \right ]^n \sum_{\{x_j\}_{j=0}^n \in S_n^{(\Lambda)}(y,x)} \Ev{\abs{G^{(\Lambda \setminus \{x_0,\ldots,x_n\})}(y,y)}^s} \\ + \left [ \frac{1}{\lambda^s (1-s)} \right ]^N \sum_{\substack{\{x_j\}_{j=0}^N \in S_N^{(\Lambda)}(x) \\
x_j\neq y,\ j=1,\ldots N}} \Ev{\abs{G^{(\Lambda \setminus \{x_0,\ldots,x_N\})}(x_N,y)}^s} 
\end{multline*}
where $$S_n^{(\Lambda)}(y,x)= \set{\text{self avoiding walks in $\Lambda$ of length $n$ from $x$ to $y$}}$$
and
$$S_N^{(\Lambda)}(x)=\bigcup_{y\in \Lambda}S_N^{(\Lambda)}(y,x) = \set{\text{self avoiding walks in $\Lambda$ of length $N$ starting at $x$}}.$$
Applying \eqref{eq:apriori} and the uniform bound $|G(x,y)|\le \nicefrac{1}{|\Im z|}$, we obtain
\begin{equation}\label{eq:bigone}
\mathbb{E}\left(\left|G^{(\Lambda)}(x,y)\right|^{s}\right) \le \sum_{n=0}^N \Gamma(s)^{1+n} \# S_n^{(\Lambda)}(y,x) + \Gamma(s)^N \# S_N^{(\Lambda)}(x) \frac{1}{|\Im z|^s}
\end{equation}
where 
\[
\Gamma \left(s\right)=\frac{1}{1-s}\frac{1}{\lambda^{s}}.
\]

If $\Gamma(s) < \nicefrac{1}{\mu_d}$ then
$$\Gamma(s)^N \#S_N^{(\Lambda)}(x) \le \Gamma(s)^N c_N \rightarrow 0$$
since the susceptibility $\chi(\Gamma(s))=\sum_n \Gamma(s)^n c_n <\infty$.
Hence taking $N\rightarrow \infty$ in \eqref{eq:bigone} we find
\begin{equation}
\mathbb{E}\left(\left|G^{(\Lambda)}(x,y)\right|^{s}\right) \le \sum_{n=0}^\infty \Gamma(s)^n \# S_n^{(\Lambda)}(y,x).
\end{equation}
Taking the supremum over volumes $\Lambda$ now yields
\begin{equation}
\sup_{\Lambda \subset \Z^d} \mathbb{E}\left(\left|G^{(\Lambda)}(x,y)\right|^{s}\right)
\le C_{\Gamma(s)}(x-y). 
\end{equation}

It remains to establish for which $\lambda$ we may find $s\in (0,1)$ with $ \Gamma(s)<\nicefrac{1}{\mu_d}$.  As above for $\lambda >e$, the unique critical point is $s_{crit}=1 - \nicefrac{1}{\ln \lambda}$ and 
$$\min_{s\in (0,1)} \Gamma(s) = \Gamma(s_{crit})= \frac{e \ln \lambda}{\lambda},$$
which is less than $\nicefrac{1}{\mu_d}$ if and only if $\lambda>\lambda_{\text{And}}$. \qed

\bibliographystyle{abbrv}
\bibliography{references}

\end{document}